\documentclass[%
 reprint,
 amsmath,amssymb,
 aps,
]{revtex4-2}

\usepackage{graphicx}
\usepackage{dcolumn}
\usepackage{bm}
\usepackage{setspace}

\begin{document}

\title{Photoemission intermittency via stochastic gating in rubrene nanowires coupled to plasmonic silver nanoparticles}

\author{Moha Naeimi}
 \altaffiliation{\textsuperscript{1}Institute of physics, University of Rostock}
 \altaffiliation{\textsuperscript{2}Department of light, life and matter, University of Rostock}

 \author{Waqas Pervez}
 \altaffiliation{\textsuperscript{1}Institute of physics, University of Rostock}
 \altaffiliation{\textsuperscript{2}Department of light, life and matter, University of Rostock}

 \author{Frithjof Harmsen}
 \altaffiliation{\textsuperscript{1}Institute of physics, University of Rostock}
 \altaffiliation{\textsuperscript{2}Department  of light, life and matter, University of Rostock}
 
\author{Ingo Barke}
 \altaffiliation{\textsuperscript{1}Institute of physics, University of Rostock}
 \altaffiliation{\textsuperscript{2}Department of light, life and matter, University of Rostock}

\author{Sylvia Speller}
 \altaffiliation{\textsuperscript{1}Institute of physics, University of Rostock}
 \altaffiliation{\textsuperscript{2}Department of light, life and matter, University of Rostock}

\date{\today}

\begin{abstract}


In this work, we report a new nanoscale phenomenon observed as photoemission intermittency (On-Off electron emission), manifested as stochastic bursts in electron yield at quasi-one-dimensional organic wires and silver nanoparticles interface. Energy-resolved measurements reveal that the emitted electrons carry out hybrid information, containing photoelectron yield enhancement associated with the nanoparticles and kinetic energies determined by the organic semiconductor. The intermittency results in a dynamic shift of the electron spectra correlating with the photoelectron yield. We attribute the observed behaviour to the photo-hole accumulation and stochastic gating of charge due to electron-hole separation at the nano interface. These findings introduces the photoemission intermittency as a nanoscale phenomenon indicating a new dynamic regime of charge assisted emission at organic–plasmonic interfaces.

\end{abstract}

\maketitle

\section{Introduction}

Organic molecular semiconductors provide a promising platform for investigating fundamental charge and exciton transport. Among them, rubrene (5,6,11,12-tetraphenyltetracene) has emerged as a high-performance molecule due to its high crystalline quality and recorded to have high charge carrier mobility in single crystalline phase, exceeding 20 $cm^2 V^{-1} s^{-1}$ \cite{Haas2007} in organic field-effect transistors, establishing a benchmark for organic electronics \cite{Podzorov2004, Podzorov2004_2, Sirringhaus2014}. Beyond charge transport, rubrene exhibits strong photoluminescence and well-resolved excitonic features \cite{Irkhin2012,Naeimi2025}, making it an ideal system to study the interplay between molecular packing and orientation, electronic structures \cite{Nitta2019}, and excited-state dynamics \cite{Clark2010}. The surface energetics, work function and ionization energy of rubrene result in band bending and charge redistribution at metal interfaces \cite{Braun2009, Ding2009}. 

In our previous work, we reported the growth of long and thin quasi-one-dimensional (quasi-1D) wires based on rubrene \cite{Naeimi2025}. One-dimensional organic crystals have emerged as a key platform in organic electronics and photonics due to their unique structural anisotropy and strong molecular coupling along a certain direction. These systems are typically formed from $\pi$-conjugated molecules (like rubrene) assembled through their $\pi$-$\pi$ stacking, enabling highly ordered molecular packing. The $\pi$-$\pi$ stacking lattice direction most often coincides with the fast growth direction and with the long axis of rubrene rods. On the other hand, inorganic nanoparticles (NPs) and hybrid organic–inorganic nanostructures are being used to tailor optical properties by exciton coupling and plasmon–exciton interactions.

In the present work, we report the observation of photoemission intermittency in rubrene wires coupled with silver (Ag) NPs grown on silicon substrate with a native $\text{SiO}_2$ layer. We combine photo-induced surface potential mapping, and photoelectron spectromicroscopy to study the dynamic gating of locally injected excitons and generated charge carriers from interfaces of rubrene wire and silver NP. Enhanced local and non-local photoluminescence was previously reported for rubrene nanowires and gold (Au) NPs due to the energy-transfer from the surface plasmon coupling \cite{Hwang2016} via polaritons. We also report an enhancement of local and non-local photoemission of rubrene wires coupled to silver NPs resulting in stochastically photoemission bursts indicating wire-NP interfaces act as localized trapped states where the mobile charge carriers can be transiently accumulated or trapped.

\section{Experimental section}

\textbf{Sample preparation and cluster deposition:} Rubrene wires with aspect ratios ranging between 20 to 300 are prepared on Si(100) substrate with native $\text{SiO}_2$ layer. The preparation method is explained in our previous work \cite{Naeimi2025}. After preparation, the sample is introduced into ultra high vacuum (UHV) environment with pressure of $\approx 10^{-10}$~mbar and was heated for 2~hours at 180~$^{\circ}\text{C}$. Afterwards the size-selected silver (Ag) NPs were deposited on the wires. We employ a magnetron sputtering source \cite{Hartmann2012} to generate silver clusters in the gas phase, a large fraction of which are singly charged. These clusters are guided through ion optics into a static quadrupole, which acts as mass selector, filtering particles by mass before they reach the substrate. By adjusting source and selector parameters, the setup can deliver beams of size-selected clusters ($\approx$ 15 nm) onto surfaces with narrow size distribution. After deposition, the NPs maintain nearly spherical shape and crystalline structure, as confirmed by AFM and TEM characterization \cite{Oldenburg2025}.

\textbf{Photoemission electron microscopy (PEEM)}: After the deposition of NPs, time-of-flight photoemission electron spectroscopy was conducted in a PEEM (Focus IS-PEEM) within the same vacuum chamber. The light angle of incidence was 23 degrees grazing. The 2nd harmonic of a tunable Ti:Sa femtosecond (fs) laser (Mira 900F) yielding photons with energies of 3.1~eV (400~nm) with the repetition rate of 1~MHz and a pulse duration of 150~ps was used as illumination source.

\textbf{Kelvin probe force microscopy (KPFM):} We used an atomic force microscope (Park Systems NX20) in sideband Kelvin probe force microscopy (KPFM) mode with conductive tips made of chromium platinum (Cr-Pt) exhibiting a cantilever spring constant of 3~N/m and a free eigenfrequency of 75~kHz. A green laser with grazing incident is also integrated to the AFM setup in order to measure illumination dependent surface potentials (surface photovoltage).

\section{Results and discussion}

\begin{figure}
    \centering
    \includegraphics[width=1\linewidth]{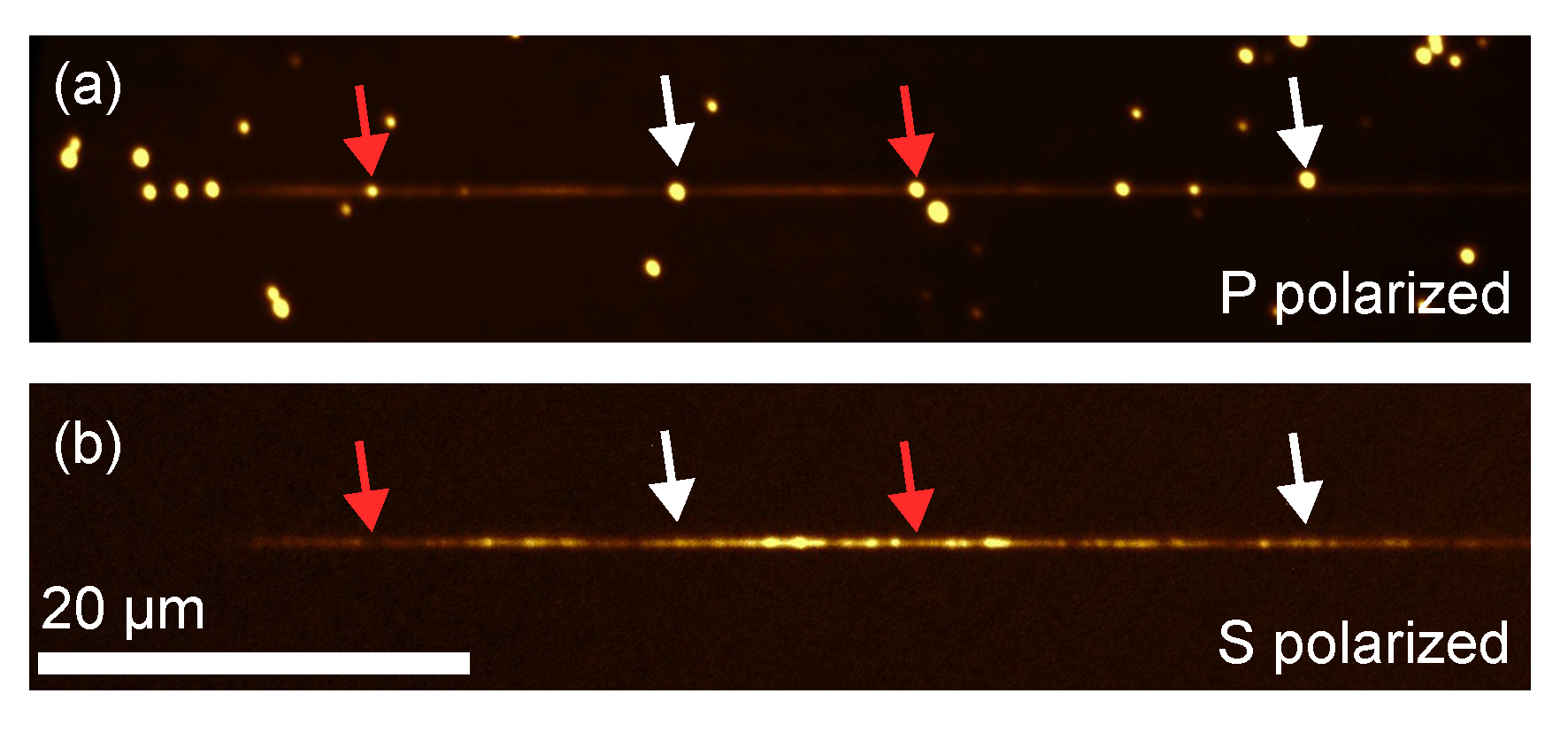}
    \caption{PEEM images of a rubrene wire decorated with silver NPs, excited by S- and P- polarized photons of 3.1 eV energy.}
    \label{AFM_fluor}
\end{figure}

Figure \ref{AFM_fluor} shows PEEM images of a rubrene wire excited by a laser with photons energies of 3.1 eV (400 nm), with S- and P- polarization. It is well visible in the PEEM images that the tendency of the deposited NPs is to attach to the wire, modifying their local photoemission and electron injection. Even though the NPs are usually non-emitting in a S-polarized excitation (Figure \ref{AFM_fluor}b), the photoemission from the wire is enhanced (white arrows in Figure \ref{AFM_fluor}a) or reduced (red arrows in Figure \ref{AFM_fluor}a) at the NP-wire interfaces or in the vicinity of them. Figure S1 shows luminescence and AFM topography map of two different rubrene wires. The photoluminescence of these wires are different and could be categorized into high-luminescence and zero-luminescence types which is not correlated with their height, indicating the different crystal orientation, in particular swap of the \textbf{b}- and \textbf{c}- axes \cite{Naeimi2025}.

The interaction of rubrene crystal assemblies with light excitation results in the generation of excitons, which are neutral quasi particles. Although any modification in surface potential indicates that the electron or holes are diffused, a signature of generation of mobile charge carriers.  Figure S2 shows the KPFM map of a rubrene wire, with and without illumination. By increasing the illumination intensity, the KPFM potential drops, indicating the generation of mobile charge carriers.

\begin{figure}
    \centering
    \includegraphics[width=1\linewidth]{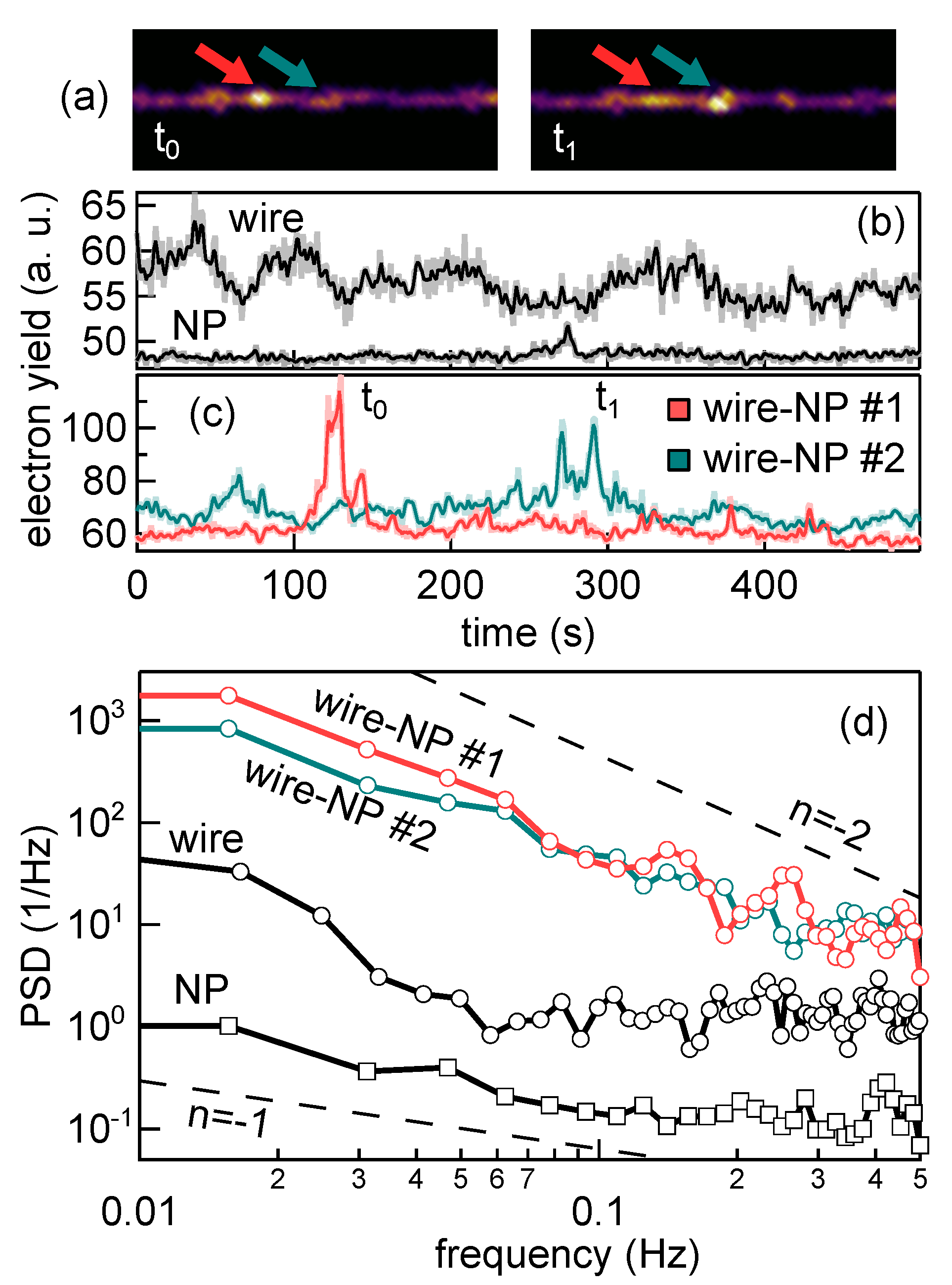}
    \caption{Photoemission intermittency is pronounced with S-polarization. (a) Two frames from a PEEM image series at two different times (t$_0$ and t$_1$), of a rubrene wire decorated with NPs, excited with S-polarized photons with energies equal to 3.1 eV. (b) Photoemission yield versus time from spots at the location of the sample inducing bare rubrene wire and isolated NP on the SiO$_2$ substrate, indicated on the plots. (c) Photoemission yield versus time from two different site with wire-NP interface, indicated in panel (a). (d) Powers spectral density of time traces in b and c.}
    \label{blinking}
\end{figure}

We observed photoemission intermittency i. e. a random switching between high and low electron emission intensity at a fixed photon excitation intensity, at the sites of wire-NP interface. The intermittency is observed as random stochastic bursts in time. Photoluminescence intermittency has been reported for molecular emitters in dyes and nanostructures such as nanocrystal, quantum dots and nanowires \cite{Galland2012}. The photoluminescence intermittency is either [A-type] in which the emission is quenched by a non-radiative recombination, or [B-type] where the hot electrons are captured by preventing them to reach the emitting states \cite{Stefani2009}. While intermittency is frequently reported for photoluminescence, to the best of our knowledge, no such effect is reported for photoelectron emission. 

It is important to emphasize that the intermittency phenomena are generally associated with fluctuations in local electronic environments and charge trapping dynamics at the nanoscale. In previously studied systems, such as quantum dots and molecular dyes, intermittency arises due to random charge trapping \cite{Cordones2013, Tang2005, Ko2011}, leading to temporal modulation of radiative recombination. In contrast, photoemission is directly the extraction of electrons into vacuum, making it sensitive not only to recombination dynamics but also to local charge variations and band bending. Therefore, the observation of intermittency in photoemission suggests a time-dependent modification of energetics in interface sites rather than purely recombination processes.

\begin{figure}
    \centering
    \includegraphics[width=1\linewidth]{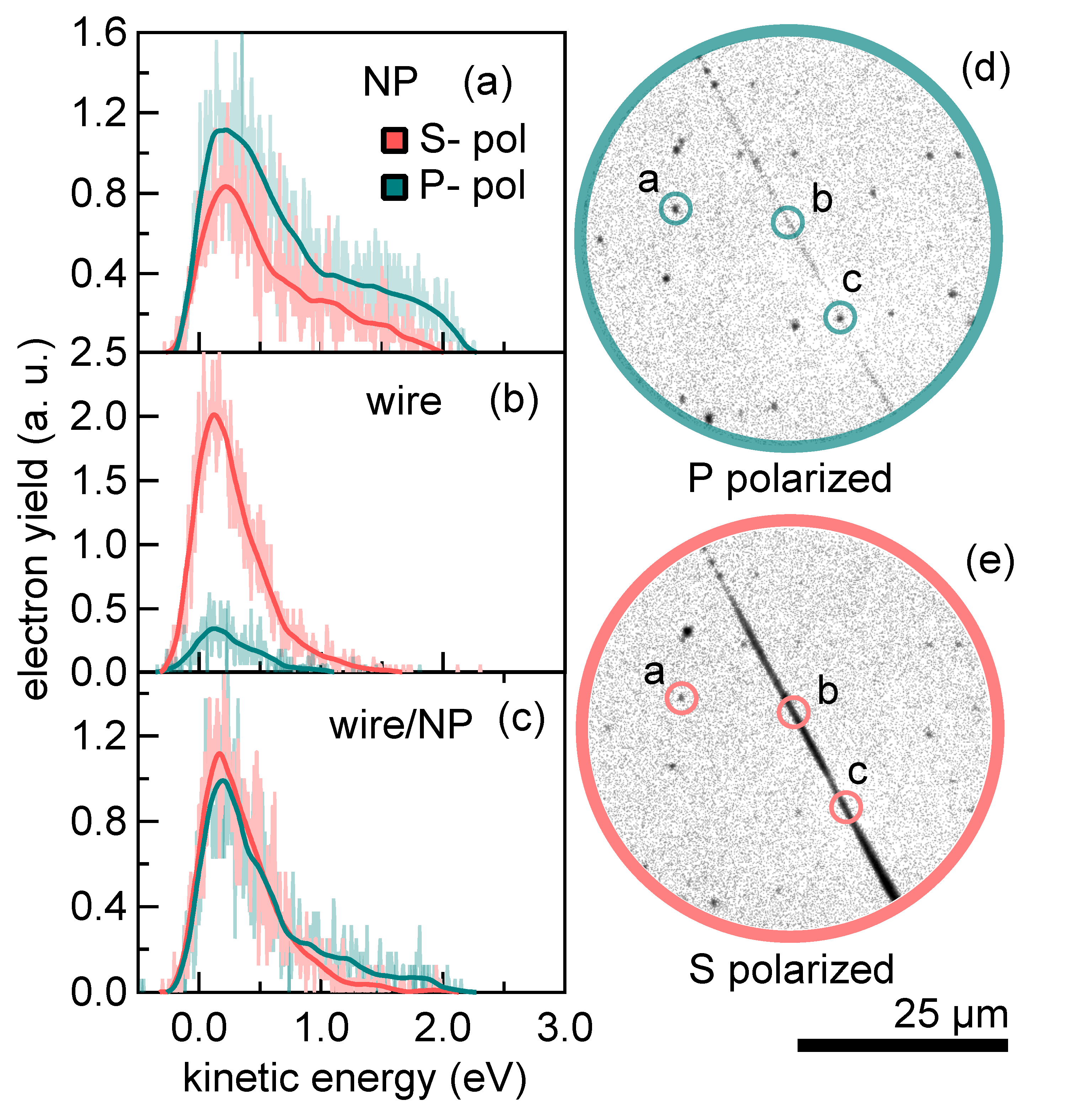}
    \caption{(a, b and c) Electron spectra of a NP, rubrene wire and wire-NP interface taken with different polarizations, respectively. (d) The respective PEEM images with marked positions related to the spectra in a, b and c.}
    \label{spec_pol}
\end{figure}

Figure \ref{blinking}a shows the photoemission yield versus time from two spots on our sample with rubrene wire and an NP, as indicated on the plot. The timetrace is a series of 500 PEEM images of a rubrene wire decorated with silver NPs (see supplementary information), each taken for one second. While a slow periodic fluctuation of photoemission is present on the wire, the NP shows constant photoemission, except a faint peak. Figure \ref{blinking} shows the photoemission yield versus time from two spots including the wire-NP interface, as indicated in the inset. Several intermittency events are observed within the time window for each interface spot. Figure \ref{blinking}c shows the power spectral density (PSD) corresponding to the time traces shown in Figures \ref{blinking}a and \ref{blinking}b. The PSD spectra from the interface sites show slopes close to -2, while the spectra from the wire show slopes lower than -2 which turns into zero for frequencies higher than 0.05 Hz. The slope of the PSD spectra from the bare NP is close to -1 and turns into zero after 0.1 Hz. Power-law behaviour in the power spectral density (PSD) has been widely reported for fluorescence intermittency in quantum dots and nanostructures, associated with charge trapping and long-lived temporal correlations \cite{Kuno2000,Pelton2004,Frantsuzov2008, Stefani2009}. PSD slopes close to -1 are commonly linked to conventional carrier dynamics in semiconductors, while steeper slopes indicate enhanced charge-memory effects and diffusion-mediated carrier dynamics. To our knowledge, such PSD behaviour has not been previously discussed in the context of photoemission intermittency.

In order to explore the origin of the photoemission intermittency, we first focus on the electron spectra at different excitation polarizations, i.e. S- and P-polarized. Then we compare the spectra at the events of high yield and low yield incidents, i. e. intermittency. Figure \ref{spec_pol}a, \ref{spec_pol}b and \ref{spec_pol}c, are electron spectra of a selected NP, a spot at the location of the rubrene wire and a spot with wire-NP interface, respectively and marked in Figure \ref{spec_pol}d and \ref{spec_pol}e. The spectra are taken with S- and P-polarized excitations, indicated by red and blue curves respectively. The wire-NP interfaces exhibit two interesting spectral features. First, the electron yield is governed by the NP, since the P-polarized excitation can merely emit electrons from the rubrene wire but the photoemission yield from the coupled-system is equal for both polarizations. Second, the spectral characteristics from the interface are inherited from the rubrene wire. These indicate that NPs enhance the photoemission from the organic wire at the interface.

\begin{figure}
    \centering
    \includegraphics[width=1\linewidth]{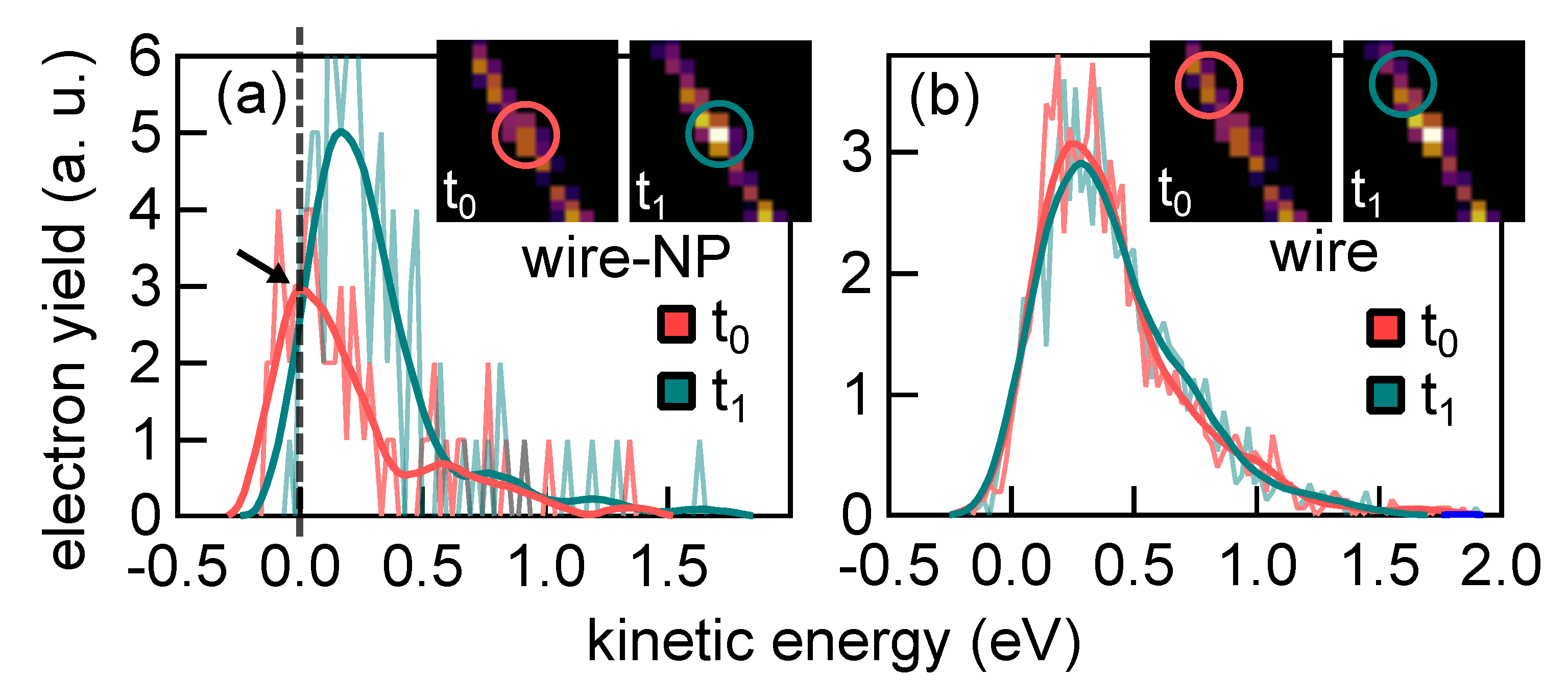}
    \caption{(a and b) Electron spectra of two sites, with and without NP attached to the wire, respectively. The spectra are resolved for highly emitting and low emitting incidents. The insets are the respective PEEM images.}
    \label{spec_blink}
\end{figure}

Figure \ref{spec_blink}a and \ref{spec_blink}b show the electron spectra at two positions with and without NP attached, respectively, from two snapshots of a PEEM image series, each taken for two seconds. The spectra from the rubrene wire- where no intermittency occurs- are similar at different times and peak at 0.5 eV. On the other hand the spectra from the wire-NP interface at different times-low yield and high yield incidents- are separated by $\sim$ 190 meV, both shifted toward lower kinetic energies. The spectral shift toward lower kinetic energies at the low yield incident is greater and consequently is more strongly bordered by the low energy cut-off, indicated by the arrow in Figure \ref{spec_blink}a. Hereby, the low photoemission yield is a result of the more invasive low energy cut-off, which means that a certain fraction of electrons can no longer overcome the work function and be emitted. The underlying mechanism of the spectral shift, stochastic gating, is discussed in the following.

The key effect is likely the transient trapping of diffusing charge carriers in vicinity of the interface between rubrene and the NP. A number of effects are expected to cooperatively contribute to the photo electron emission intermittency. At the coupling region of the nanoparticles with the rubrene lattice a zone of local disorder exists, giving rise to trap sites modifying the local density of states. The NP acts as a perturbation to the local potential, where carriers can transiently be trapped. Roaming and trapping of charge carriers leads to dynamic or stochastic gating. In addition, excitation at 400 nm is resonant with the plasmonic response of the silver NPs \cite{Oldenburg2018} resulting in strongly localized and enhanced electromagnetic near fields at the wire-NP interface. This local field enhancement is expected to amplify exciton and carrier generation. A third determinant is the role of the quasi-1D rod geometry in combination with the wire-NP interfaces. Such rods confine roaming charge carriers and can act as active waveguides for fluorescence or exciton-polariton light leading to a swift redistribution of energy along the longitudinal directions of the rod. All this together enhances the probability of observing intermittency in form of transient “hot spots” of photo-emission, induced by diffusing charge carriers and the respective stochastic gating.

We attribute the energetic shifs to hole accumulation at the wire-NP interface as metastable configurations, where trapping sites approach saturation and the local electrostatic potential modifies electron emission conditions. The "normal" state of photoemission at the interface region is in fact the high yield or On-state, since the emission is enhanced by the NPs (see Figure \ref{AFM_fluor}). The modification of local electrostatic potential by photo-hole accumulation, shifts the spectral energy of emitted electrons toward lower kinetic energies, eventually it falls below the work function and the liberation of photoelectrons gets reduced. This introduces the "abnormal" state of photoemission, i. e. low yield or off-state. As a result, under constant excitation intensity, the system can exhibit dynamics characterized by stochastic gating of electron emission \cite{Yuan2016}. 

Now we can compare the PSD slopes, i. e. the power-law scaling exponents, and relate them to the underlying intermittency mechanism. While both, rubrene wire and bare NP, exhibit PSD slopes of roughly -2 and -1, respectively, approaching the detection limits after $\approx$ 0.05 Hz, the PSD slope of the wire–NP interface sites is close to -2 and does not reach the detection limit. In 3 dimensions, a scaling exponent of -2 would indicate diffusion. The narrowness of the rod should confine the random walk of the carriers, and may give rise to scaling exponents higher than -2 \cite{Pfluegl1998}. In that case -2 would indicate already a super-diffusive mechanism. The question is whether the intermittency is encoded in the PSDs? Memory” effects have been proposed, suggesting that the gating dynamics exhibit a dependency on prior system configurations, rather than being fully stochastic in nature. This could indeed apply to a trap or interface site. Such memory effects rendering particularly the low-frequency contributions high, and thereby slopes steep \cite{Muoz2022}. The interface leads the diffusive behaviour being extended to higher frequencies, consistent with diffusion-assisted charge accumulation and stochastic discharge processes.

These findings open a perspective on the role of exciton dynamics in charge extraction processes in organic semiconductor systems. The observed correlation between emission intensity and kinetic energy suggests that intermittency is not merely an intensity fluctuation but is intrinsically linked to energy redistribution mechanisms at the interface. This could have important implications for understanding charge transfer processes in hybrid organic–nanostructured systems, particularly in devices where interfaces dominate performance, such as photodetectors and optoelectronic junctions.

\section*{Conclusion}
We reported the observation of photoemission intermittency from organic–plasmonic interfaces characterized by stochastic gating correlated with shift in the energy of photoelectrons. The observed coupling between the photoelectron yield and energy reveals a direct link between photoemission and charge accumulation at interfaces, pointing to a dynamic gating of the local potential by mobile carriers. Beyond the specific rubrene–Ag system, our findings suggest that stochastic, charge-driven emission may be a general feature of organic-metal interfaces by combining mobile charge carriers with localized trapping sites, introducing new opportunities to probe interfacial charge dynamics at the nanoscale.

\section*{Acknowledgement}

Funding by the Deutsche Forschungsgemeinschaft (DFG, German Research Foundation) within projects SFB 1477 “Light-Matter Interactions at Interfaces” (441234705), SFB 1270/2 “Electrically Active Implants” (299150580) and “Application of Interoperable Metadata Standards (AIMS) 2” (432233186) is acknowledged.

\bibliography{rsc}

\end{document}